\documentclass[12pt]{iopart}
\pdfoutput=1
\usepackage{iopams}
\expandafter\let\csname equation*\endcsname\relax

\expandafter\let\csname endequation*\endcsname\relax

\usepackage{citesort}

\usepackage{array}
\usepackage{multirow}
\usepackage{multicol}
\usepackage{ulem}
\usepackage{amsmath}
\usepackage[a4paper, total={6in, 10in}]{geometry}
\usepackage[numbers]{natbib}
\usepackage{graphicx}
\usepackage[utf8]{inputenc}
\usepackage[T1]{fontenc}
\usepackage{physics}
\usepackage[breaklinks=true,colorlinks=true,linkcolor=blue,urlcolor=blue,citecolor=blue]{hyperref}
\hypersetup{
	colorlinks   = true,
	citecolor    = cyan,
	urlcolor     = red,
}
\usepackage[usenames, dvipsnames]{color}
\usepackage[mathscr]{euscript}
\usepackage[T1]{fontenc}
\usepackage[font=small,labelfont=bf,tableposition=top]{caption}

\DeclareCaptionLabelFormat{andtable}{#1~#2  \&  \tablename~\thetable}

\begin{document}

\title[Ultracold polar molecules as qudits]{Ultracold polar molecules as qudits}

\author{Rahul~Sawant$^{1,2}$, Jacob~A~Blackmore$^1$, Philip~D~Gregory$^1$, Jordi~Mur-Petit$^3$,
Dieter~Jaksch$^{3,4}$,
Jes{\'u}s~Aldegunde$^5$,  Jeremy~M~Hutson$^6$, M~R~Tarbutt$^7$ and Simon~L~Cornish$^1$}
\address{$^1$Joint Quantum Centre (JQC) Durham-Newcastle, Department of Physics, Durham University, South Road, Durham DH1 3LE, United Kingdom.\\
$^2$ Midlands Ultracold Atom Research Centre, School of Physics and Astronomy, University of Birmingham, Edgbaston, Birmingham B152TT, United Kingdom. \\
		$^3$ Clarendon Laboratory, University of Oxford, Parks Rd, Oxford OX1 3PU, United Kingdom.\\
		$^4$ Centre for Quantum Technologies, National University of Singapore, 3 Science Drive 2, 117543 Singapore. \\
		$^5$ Departamento de Quimica Fisica, Universidad de Salamanca, 37008 Salamanca, Spain. \\
		$^6$ Joint Quantum Centre (JQC) Durham-Newcastle, Department of Chemistry, Durham University, South Road, Durham DH1 3LE, United Kingdom.\\
		$^7$ Centre for Cold Matter, Blackett Laboratory, Imperial College London, Prince Consort Road, London SW7 2AZ, United Kingdom.\\
		}
\eads{\mailto{R.V.Sawant@bham.ac.uk}, \mailto{s.l.cornish@durham.ac.uk}}

\begin{abstract}
We discuss how the internal structure of ultracold molecules, trapped in the motional ground state of optical tweezers, can be used to implement qudits. 
We explore the rotational, fine and hyperfine structure of $^{40}$Ca$^{19}$F and $^{87}$Rb$^{133}$Cs, which are examples of molecules with $^2\Sigma$ and $^1\Sigma$ electronic ground states, respectively. In each case we identify a subset of levels within a single rotational manifold suitable to implement a 4-level qudit. Quantum gates can be implemented using two-photon microwave transitions via levels in a neighboring rotational manifold. We discuss limitations to the usefulness of molecular qudits, arising from off-resonant excitation and decoherence. As an example, we present a protocol for using a molecular qudit of dimension $d=4$ to perform the Deutsch algorithm.  
\end{abstract}
\vspace{2pc}

\maketitle

\noindent
Quantum computation has the potential to outperform conventional computation for certain challenging  problems~\cite{Montanaro2016}. Many groups are developing the building blocks of a quantum computer, exploring several different physical systems in the search for the best architecture~\cite{Imamoglu1999,Vandersypen2001,Kielpinski2002,Knill2001,Neumann2008,Clarke2008,Saffman2010,debnath_demonstration_2016,saffman_quantum_2016,Zheng_2017,watson_programmable_2018,qiang_largescale_2018,borders_integer_2019}.  
One of the challenges is the problem of scalability~\cite{divincenzo_physical_2000}: it is difficult to engineer a quantum system with a large number of individually controllable qubits that together form a large Hilbert space and are free from external perturbations and loss.     

\begin{figure}[t]
	\includegraphics[width=12cm]{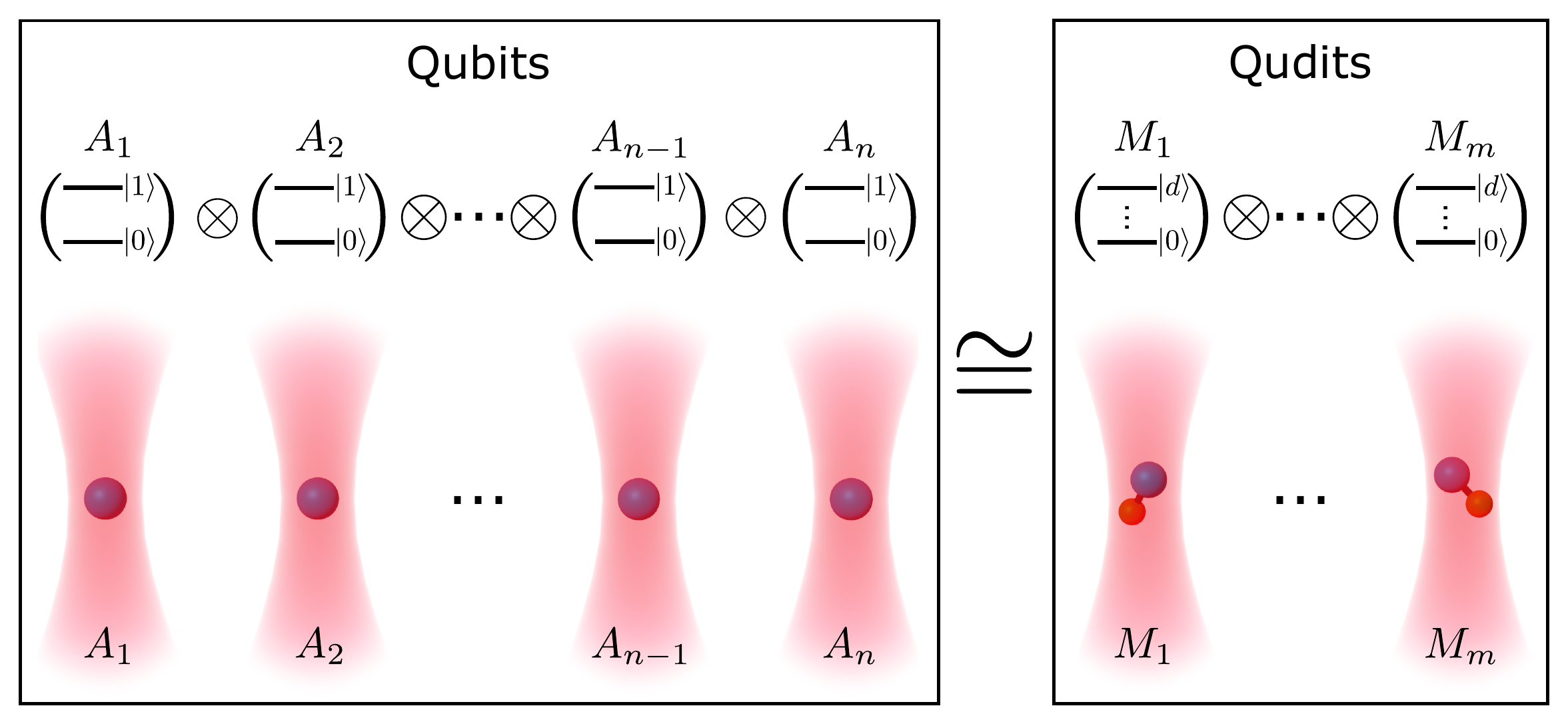}
	\centering
	\caption{Combining qubits and qudits formed in optical tweezer arrays. $p$ qubits form a Hilbert space with dimension $2^p$, while $q$ qudits form a space of dimension $d^q$. To achieve the same Hilbert-space dimension in each case, $q = p/\log_2 d$. This shows the advantage of using qudits for quantum computation.}
	\label{equivalence}
\end{figure}

The problem of scalability can be reduced by using higher-dimensional quantum systems (qudits) instead of two-level qubits. 
For the same size of Hilbert space, the number of $d$-level qudits required is smaller than the number of qubits by the factor $\log_2 d$~\cite{Muthukrishnan2000,Bartlett2002}, as shown in figure~\ref{equivalence}. 
For example, to perform a computation that is beyond the capabilities of any current classical computer (termed quantum supremacy~\cite{preskill_quantum_2012}) requires about 50 qubits~\cite{Neill2018} but only 15 10-level qudits.
Additionally, the time required to carry out gate operations can be reduced by a factor of $(\log_2 d)^2$~\cite{Muthukrishnan2000,stroud_quantum_2002} if arbitrary transformations can be achieved in the $d$-dimensional space. Other advantages of using qudits for quantum computation are increased robustness~\cite{Zilic2007,Parasa2011,Parasa2012} and improvements for quantum error-correcting codes~\cite{Campbell2012,Duclos-Cianci2013,anwar_fast_2014,Campbell2014,Andrist2015,Watson2015,krishna_2019}. 

There are many quantum-computational algorithms that can work with even a unary (single) qudit. An important example is Grover's search algorithm~\cite{Lloyd1999, Ivanov2012}. Versions of this have been implemented using an optical field as a qudit with $d=4$~\cite{kwiat_grovers_2000}, 
atoms with $d=8$~\cite{ahn_information_2000} and a single nuclear spin with $d=4$~\cite{Godfrin2017}. Other algorithms that can be performed with qudits include quantum phase estimation and quantum counting~\cite{Tonchev2016}, the Deutsch algorithm~\cite{Kiktenko2015}, and finding the parity of permutation; the last has been experimentally demonstrated by Gedik \textit{et al.}~\cite{Gedik2015} with a single qudit, using nuclear magnetic resonance. 

Ultracold molecules provide very attractive systems for quantum computation. The rotational and spin degrees of freedom make it possible to encode quantum information in ways not possible on other platforms. Experiments with ultracold molecules have progressed rapidly over the last decade~\cite{Ni2008,Danzl:2008,Lang:2008,Takekoshi2014,Molony2014,Park:2015,Seeselberg2018,Guo:2016,Rvachov:2017,liu_building_2018,Liu_2019,barry_magneto-optical_2014,truppe_molecules_2017,anderegg_laser_2018,Collopy_2018}, and the rotational, fine and hyperfine structure has been studied in detail \cite{aldegunde_hyperfine_2008,aldegunde_microwave_2009, ran_microwave_2010,Neyenhuis_2012, Will2016, Gregory2016, aldegunde_hyperfine_2017,Aldegunde2018}. Heteronuclear molecules can have electric dipole moments fixed in the molecular frame, allowing manipulation of the quantum states with microwave fields~\cite{Neyenhuis_2012, Will2016, Gregory2016, Park2017}. 
A quantum computer formed from ultracold polar molecules can be increased in scale by linking neighboring molecules via the long-range dipole-dipole interaction~\cite{DeMille2002,Yelin_2006,Zhu_2013,Herrera_2014,Karra_2016,Ni2018,hughes_robust_2019}. 

In this article, we investigate the rich internal structure of diatomic molecules with the aim of using them as qudits. We consider ultracold molecules trapped in the motional ground states of optical tweezers. Tweezers are an established tool for atoms~\cite{barredo2016,endres2016,Norcia2018,Cooper2018,barredo_synthetic_2018,Mello_2019}, and have recently been extended to ultracold molecules, both by loading laser-cooled molecules directly into tweezers~\cite{anderegg2019} and by associating atoms in a tweezer to form molecules~\cite{liu_building_2018, Liu_2019}. 
Tweezers offer single-particle addressability and detection, combined with easy scaling up to arrays of $\sim 100$ traps~\cite{Cooper2018,barredo_synthetic_2018,Mello_2019}, making them an ideal platform for quantum computation with ultracold molecules.

In the following sections, we examine the rotational and hyperfine structure of ultracold molecules in electronic states of $^{2}\Sigma$ and $^{1}\Sigma$ symmetry, with a focus on $^{40}$Ca$^{19}$F~\cite{truppe_molecules_2017} and $^{87}$Rb$^{133}$Cs~\cite{Molony2014}. In each case we identify a set of levels that can form a qudit, with transitions between them well isolated from others that might cause loss. In section~\ref{decoherence_sec}, we discuss the sources of decoherence of these qudits and the limitations these may present in future experiments.  We focus on two main sources of decoherence: differential ac-Stark shifts and magnetic field noise. In section~\ref{gate_sec}, we discuss basic gate operations for the qudits, implementable using microwaves. Finally, in section~\ref{Deutsch_section}, we describe how the Deutsch algorithm~\cite{Deutsch_david_quantum_1985} can be implemented using a single qudit formed from a single diatomic molecule.

\section{Internal structure of ultracold molecules relevant for qudits}
The internal structure of molecules is very rich, even in the electronic and vibrational ground state, because of the presence of molecular rotation, electron spin and nuclear spins. In this section, we briefly discuss the advantages and challenges this presents for the realization of qudits. We describe the internal structure of diatomic molecules in $^2\Sigma$ and $^1\Sigma$ electronic states. Molecules in $^2\Sigma$ states possess an unpaired electron, whereas those in $^1\Sigma$ states do not. $^1\Sigma$ is the electronic ground state of molecules formed by associating two alkali atoms~\cite{Ni2008,Danzl:2008,Lang:2008,Takekoshi2014,Molony2014,Park:2015,Seeselberg2018,Guo:2016,Rvachov:2017,liu_building_2018, Liu_2019}.
$^2\Sigma$ is the electronic ground state of molecules such as SrF~\cite{barry_magneto-optical_2014}, CaF~\cite{truppe_molecules_2017,anderegg_laser_2018}, YbF~\cite{Lim_2018} and YO~\cite{Collopy_2018},  which have been recently laser cooled. Molecules formed by associating an alkali atom and a closed-shell atom~\cite{Munchow:2011, barbe_observation_2018,Guttridge2018} will also have $^2\Sigma$ ground states.  
To illustrate our discussion, we consider the specific cases of $^{40}$Ca$^{19}$F and $^{87}$Rb$^{133}$Cs molecules.
\subsection{General considerations}
We propose to use the rotational and hyperfine levels of molecules to form a qudit. A qudit of dimension $d$ is formed from $d$ primary levels in a single rotational manifold and can be manipulated using two-photon microwave transitions via auxiliary levels in a neighbouring rotational manifold.
The transitions must be sufficiently separated in frequency to avoid off-resonant excitation to other levels (both within the qudit and, more broadly, within the molecule). 
Any such excitation will reduce the state fidelity.
The upper bound for the probability of an off-resonant excitation is~\cite{steck_quantum_2011,MurPetit2012},
\begin{equation}
\label{loss_frac}
p_\text{loss} = \frac{(r_\textrm{tdm}\Omega)^2}{(r_\textrm{tdm}\Omega)^2+\Delta^2},
\end{equation}
where $\Omega$ is the Rabi frequency for the target transition, $\Delta$ is the frequency detuning of the microwave field from the off-resonant transition, and $r_\textrm{tdm}$ is the ratio of the transition dipole moments of the off-resonant and target transitions. $\Omega$ is related to the duration $t_{\pi/2}$ of a $\pi/2$ pulse by $\Omega =  \pi/(2 t_{\pi/2})$. 

We consider molecules trapped in optical tweezers with wavelength $\lambda  = 1064$ nm and waist 1~$\mu$m. The tweezers are assumed to have radial and axial trapping frequencies $\omega_r \sim 10$~kHz and $\omega_z \sim 2$~kHz, respectively. This is achieved with a peak laser intensity of $I_0 \sim 20$ kW cm$^{-2}$ for $^{40}$Ca$^{19}$F and $I_0 \sim 5$ kW cm$^{-2}$ for $^{87}$Rb$^{133}$Cs.~\footnote{In calculating the intensity, we use values of the polarizability at 1064~nm for $^{87}$Rb$^{133}$Cs from experiment~\cite{blackmore_unpublished} and for $^{40}$Ca$^{19}$F from theory~\cite{Tarbutt_unpublished}.}
Our choice of $I_0$ is a compromise between two considerations.
First, as explained in section~\ref{decoherence_sec}, the noise in the intensity of tweezers leads to decoherence, which is in general more severe at high intensities. However, lowering the intensity increases the width of the external wavefunction of the molecule trapped in the tweezer~\cite{Ni2018}. This limits the proximity achievable before molecules can tunnel between tweezers and thus reduces the achievable dipole-dipole interactions. 
\subsection{$^{2} \Sigma$ diatomic molecules}\label{2sigma}
\begin{figure}[h]
	\includegraphics[width=11cm]{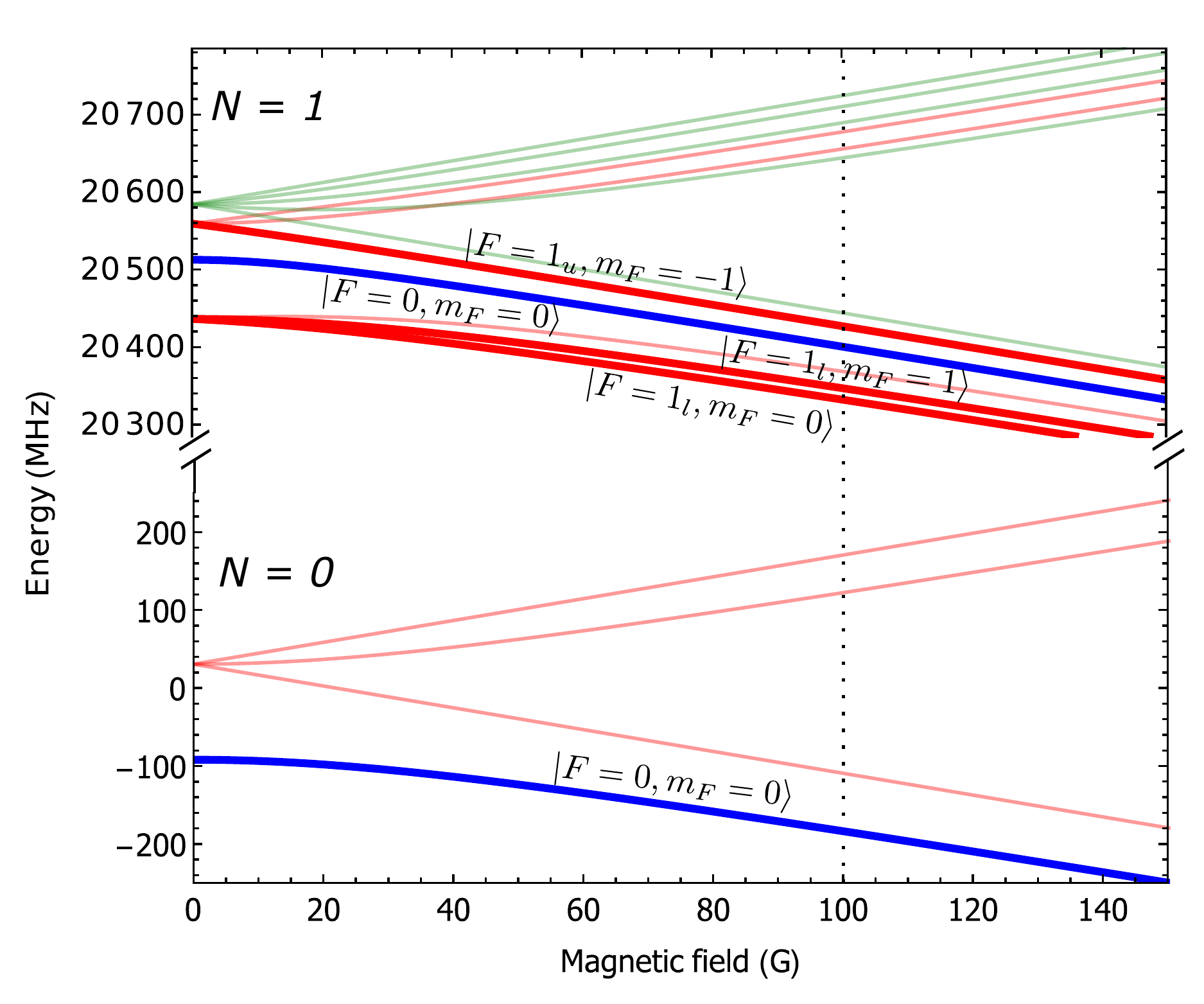}
	\centering
	\caption{Zeeman splitting of hyperfine levels of $^{40}$Ca$^{19}$F. 
	The colors blue, red, and green indicate $F$ = 0, 1 and 2 respectively. The thick lines represent the levels selected as described in section~\ref{2sigma}. The vertical dotted line shows the magnetic field $B = 100$~G where we propose to use the molecule as a qudit.}
	\label{CaF_levels}
\end{figure}
We start by discussing the case of a $^{2} \Sigma$ molecule. The Hamiltonian of such a molecule in the electronic and  vibrational ground state has been described in refs.~\cite{Aldegunde2018,Brown:2003}.
The number of hyperfine levels in rotational manifold $N$ is $(2N+1)(2S+1)(2I_1 +1)(2I_2 + 1)$, where $S$ and $I_i$ are the total electron spin and the spins of the two nuclei, respectively. $N$, $S$ and $I_i$ couple to form a resultant $F$, with projection $m_F$ on the magnetic field axis. $F$ is a good quantum number at zero field, but not at finite field. With $S = 1/2$, $I_1 = 0$ and $I_2 = 1/2$, $^{40}$Ca$^{19}$F has 4 hyperfine levels in $N = 0$ and 12 in $N = 1$, as shown in figure~\ref{CaF_levels}. Each level is identified by ($N,F_\textrm{l/u},m_F)$, where the subscripts l and u specify the lower and upper levels with $F=1$. 

The difference between the hyperfine splittings of the $N$ and $N \pm 1$ rotational manifolds depends largely on two terms: (i) the interaction between the electron spin and the nuclear rotation and (ii) the dipolar interaction between the electron and nuclear spins~\cite{blackmore_ultracold_2018}.   
For $^{40}$Ca$^{19}$F, the contributions of these two terms to the splitting are $\sim 60$ MHz and $\sim 20$ MHz respectively~\cite{Aldegunde2018}. In contrast, the ac-Stark effect shifts the energy levels by less than 50~kHz relative to each other and may be neglected.

The four hyperfine levels $e$ shown in Table~\ref{CaF_4levels} can be used to define a 4-level qudit space with $N=1$. The level $g$ with $(N,F,m_F)=(0,0,0)$ has transitions to each of the selected $N = 1$ levels in a magnetic field and functions as the auxiliary level. At a magnetic field of 100~G, the transitions are sufficiently isolated and have transition dipole moments $\mu_{eg}$ greater than $0.05$~D. For a reasonably fast microwave pulse, $t_{\pi/2} = 5 \ \mu$s, a frequency detuning greater than 10 MHz results in minimal loss, $p_\text{loss} < 10^{-5}$ (using equation~\eqref{loss_frac}), where we assume $r_\textrm{tdm} = 1$. 
\begin{table}[h]
\begin{center}
 \begin{tabular}{||p{1.7cm}|p{1.3cm}|p{2cm}|p{2.3cm}||}
    \hline
    $e$  & $g$   & $f$/MHz & $\mu_{eg}$/D\\
    \hline
    $(1,1_\text{l},1)$ & $(0,0,0)$  & 20515.969  & 1.76 \\
    $(1,1_\text{l},0)$ & $(0,0,0)$  & 20530.739  & 1.76 \\
    $(1,0,0)$ & $(0,0,0)$   & 20584.305  & 0.07\\
    $(1,1_\text{u},-1)$ & $(0,0,0)$  & 20610.756  & 0.27 \\
    \hline
    \end{tabular}
\end{center}
\caption{Levels $e=(N,F_\textrm{l/u},m_F)$ that form a 4-level qudit for $^{40}$Ca$^{19}$F with $N=1$, together with the isolated transitions to the level $g=(0,0,0)$ that can be used to manipulate them at a magnetic field of 100~G and tweezer intensity of 20 kW cm$^{-2}$.} 
\label{CaF_4levels}
\end{table}

If qudits of higher dimension are required, hyperfine levels of rotational manifolds with $N > 1$ can be used. $^2\Sigma$ molecules with higher nuclear spins, such as $^{87}$Rb$^{87}$Sr and $^{133}$Cs$^{173}$Yb, will also allow the formation of higher-dimensional qudits, even in their $N = 0$ rotational manifolds. There are 80 and 48 hyperfine levels in the $N = 0$ manifolds for $^{87}$Rb$^{87}$Sr~\cite{Aldegunde2018} and $^{133}$Cs$^{173}$Yb~\cite{Guttridge2018}, respectively. 
This illustrates a significant advantage of ultracold molecules over atoms: simple diatomic molecules have substantially more levels in their electronic ground states than atoms~\cite{Chaudhury2007,chaudhury_quantum_2009,Smith2013,Anderson2015}. 

\subsection{$^{1}\Sigma$ diatomic molecules}\label{1sigma}
The Hamiltonian of a $^1\Sigma$ diatomic molecule in its electronic and vibrational ground state is discussed in refs.~\cite{aldegunde_hyperfine_2008,aldegunde_hyperfine_2017}. The difference between the hyperfine splittings of the $N$ and $N \pm 1$ rotational manifolds depends on the interaction of the nuclear spins and molecular rotation, principally through nuclear electric quadrupole terms.
For $^{87}$Rb$^{133}$Cs molecules the quadrupole terms contribute around 0.5 MHz for $N=1$, although there are other alkali dimers such as $^6$Li$^{85}$Rb for which the contributions are an order of magnitude larger \cite{aldegunde_hyperfine_2017}. 
The splittings are two orders of magnitude smaller for $^{87}$Rb$^{133}$Cs than for $^{40}$Ca$^{19}$F.

The eigenstates may be expanded in an uncoupled basis set $\ket{N,m_N,m_I^\text{Rb},m_I^\text{Cs}}$ where $N$ is the rotational quantum number and $m_N,m_I^\text{Rb}$ and $m_I^\text{Cs}$ are projections of $N$ and the nuclear spins along the magnetic field. 
The projection quantum numbers are conserved at sufficiently high magnetic fields, but the hyperfine part of the Hamiltonian is off-diagonal and introduces mixings. Transitions between many hyperfine levels are allowed \cite{aldegunde_microwave_2009, ran_microwave_2010}. 
We calculate transition dipole moments between pairs of eigenstates as described in refs.\ \cite{aldegunde_microwave_2009, ran_microwave_2010} and define the transition strength as the square of the transition dipole.

The number of hyperfine levels in rotational manifold $N$ is $(2N+1)(2I_1 +1)(2I_2 + 1)$. 
For $^{87}$Rb$^{133}$Cs, with $I_1 = 3/2$ and $I_2 =  7/2$, this gives 32 levels for $N = 0$ and 96 levels for $N = 1$; these have complicated hyperfine and Zeeman splittings, as shown in Figs.~\ref{RbCs_levels_N0}(a) and \ref{RbCs_levels_N1}(a). They are further split and shifted by ac-Stark effects that are comparable to the hyperfine splittings, as shown in Figs.~\ref{RbCs_levels_N0}(b) and \ref{RbCs_levels_N1}(b) for laser polarization parallel to the magnetic field. In this proposal, the magnetic field is held at 181.5 G, which is where the molecules are created in current experiments~\cite{Takekoshi2014, Molony2014}. Despite the close level spacing at this field, we can still find transitions that are sufficiently isolated from each other. The spacing between the hyperfine levels can be increased by increasing the magnetic field, but this decreases the mixing of the uncoupled spin states, leading to reduced transition strengths to some states. The higher microwave power required then increases the probability of unwanted transitions. 
 
To identify levels that can form a qudit, we consider the eigenstates of the Hamiltonian of  Ref.~\cite{Gregory2017}, which includes rotational, hyperfine, Zeeman, and ac-Stark effects. 
All the spectroscopic constants for this Hamiltonian have been determined by microwave spectroscopy~\cite{Gregory2016,Gregory2017}. We use a laser intensity of $5$ kW cm$^{-2}$ and polarization parallel to the magnetic field. The following procedure is used to find as many connected levels as possible with transitions that are isolated for chosen values of $t_{\pi/2}$ and $p^\textrm{max}_\textrm{loss}$. 
\begin{enumerate}
    \item Start with the lowest level in the $N=0$ (or $N=1$) rotational manifold. 
    \item Select an additional primary level with the same $N$ (initially the next one in energy) for consideration to add to the qudit.
    \item Consider each combination of a level already in the qudit and a level in the auxiliary manifold for a two-photon transition to the new primary level. Reject the combination if either of the transition strengths $|\mu_{eg}|^2$ is less than $0.01d_0^2/3$, where $d_0$ is the permanent electric dipole moment of the molecule.
    \item At each frequency required for the augmented set of transitions, calculate the off-resonant excitation probability, $p_\text{loss}$, from each level in the qudit to every unwanted level in the auxiliary manifold. If $p_\textrm{loss} > p^\textrm{max}_\textrm{loss}$, reject this combination. 
    \item If $p_\textrm{loss} > p^\textrm{max}_\textrm{loss}$ for all possible combinations, reject the candidate primary level. Return to (ii) to consider a new candidate.
    \item If $p_\textrm{loss} < p^\textrm{max}_\textrm{loss}$ for one or more combinations, add the new primary level to the qudit, with the corresponding auxiliary level and transitions. If possible choose an auxiliary level that is already in the set; to distinguish between candidates, choose the one with the highest product of transition strengths. Return to (ii) to consider adding an additional primary level. 
\end{enumerate}
In calculating $p_\text{loss}$, we assume that the polarization of the microwave has $95 \ \%$ purity. For example, if we want to drive a $\sigma_+$ transition at frequency $f$ and Rabi frequency $\Omega$, the $\sigma_-$ and $\pi$ transitions will be driven by frequency $f$ and Rabi frequency $\sqrt{0.05}\ r_\textrm{tdm} \Omega$. The calculations assume square pulses for population transfers; shaped pulses can in principle reduce leakage by one further order of magnitude~\cite{Motzoi2009}, but we have not investigated this in detail. 

The results of such searches for $N = 0$ and $N = 1$ rotational manifolds of $^{87}$Rb$^{133}$Cs are discussed below. 
\subsubsection{Qudit with $N = 0$ levels.}
\begin{figure}[b]
	\includegraphics[width=15cm]{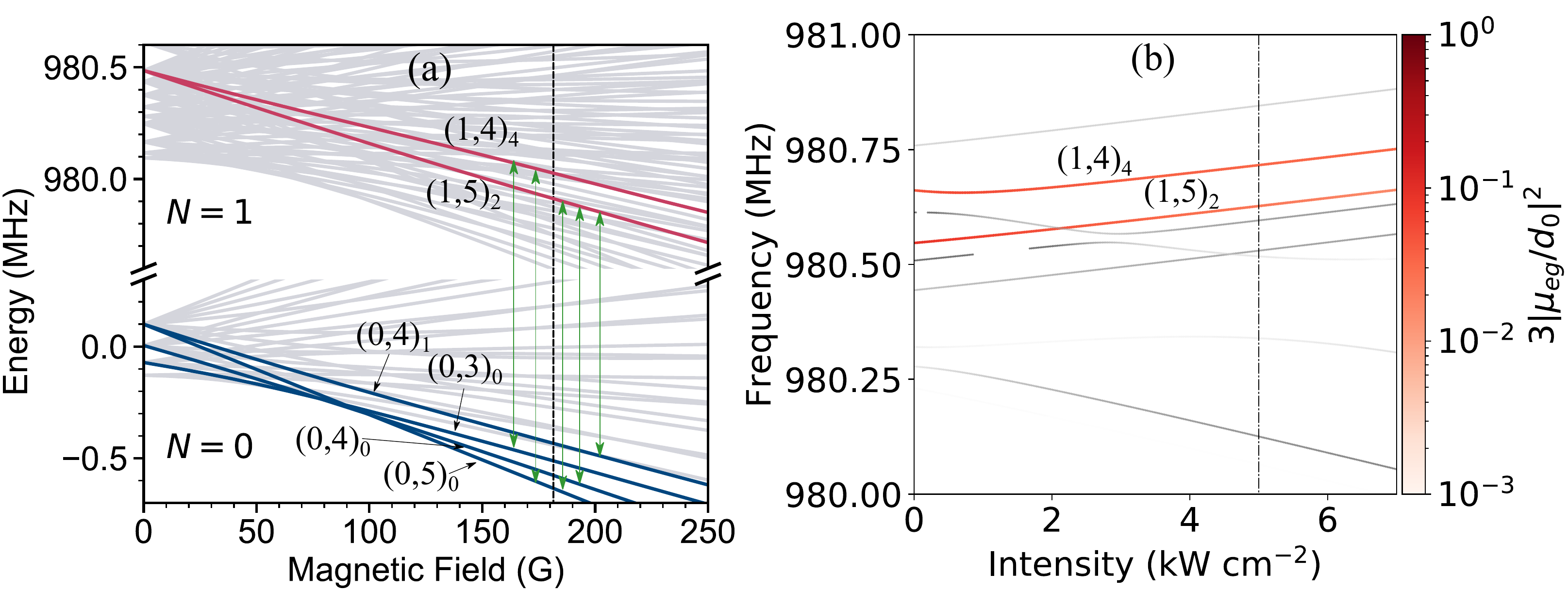}
	\centering
	\caption{(a) Zeeman splitting of hyperfine levels of $^{87}$Rb$^{133}$Cs. The vertical dashed line indicates a magnetic field $B=181.5$ G. Levels that form a 4-level qudit with $N=0$ are highlighted in blue. Levels in $N = 1$ that can be used to manipulate them are highlighted in red, with the transitions shown by the green arrows (separated in magnetic field for clarity). (b) Frequencies of transitions from level $(0,5)_0$, including the ac-Stark shift. Intensities are shown by shading. The dot-dashed line indicates an intensity $I_0=5$ kW cm$^{2}$. The ac-Stark shift is the same for all levels with $N=0$.} 
	\label{RbCs_levels_N0}
\end{figure}
The levels of the $N = 0$ manifold have equal ac-Stark shifts~\cite{Gregory2017}, so are highly attractive as potential qudits. The transfer between the $N = 0$ levels can be achieved via auxiliary levels with $N = 1$. We use the algorithm described above to search for hyperfine levels with isolated transitions. For $t_{\pi/2} = 1$ ms and $p^\textrm{max}_\textrm{loss} = 3\times10^{-3}$ we find 8 hyperfine levels in the $N = 0$ manifold that are connected with one another via at least one common $N = 1$ level. For a shorter pulse duration of $t_{\pi/2} = 0.3$ ms, we find 4 such levels. These are listed in table~\ref{rbcs_n0_levels}, together with the isolated transitions that can be used to manipulate them. The levels are highlighted in figure~\ref{RbCs_levels_N0}. The states are labelled $(N,m_F)_i$, where the subscript $i$ distinguishes between levels that have the same values of $N$ and $m_F$ but differ in energy; the lowest level for each $(N,m_F)$ is labelled $i=0$. 
\begin{table}[h]
\begin{center}
    \begin{tabular}{||p{1.1cm}|p{1.1cm}|p{2cm}|p{1.8cm}||}
    \hline
    $g$  & $e$   & $f$/kHz & $3|\mu_{eg}|^2/d_0^2$\\
    \hline
    $(0,4)_1$ & $(1,5)_2$  & 980426.03  & 0.93 \\
    $(0,4)_0$ & $(1,5)_2$  & 980569.25  & 0.04 \\
    $(0,3)_0$ & $(1,4)_4$  & 980593.00  & 0.03 \\
    $(0,5)_0$ & $(1,5)_2$  & 980627.29  & 0.04 \\
    $(0,5)_0$ & $(1,4)_4$  & 980716.34  & 0.03 \\
    \hline
    \end{tabular}
\end{center}
\caption{Levels $g=(N,m_F)_i$ that form a 4-level qudit for $^{87}$Rb$^{133}$Cs with $N=0$, together with the isolated transitions to levels with $N=1$ that can be used to manipulate them at a magnetic field of 181.5~G and tweezer intensity of 5 kW cm$^{-2}$.}
\label{rbcs_n0_levels}
\end{table}

\subsubsection{Qudit with $N = 1$ levels.} 
\begin{figure}[b]
	\includegraphics[width=15cm]{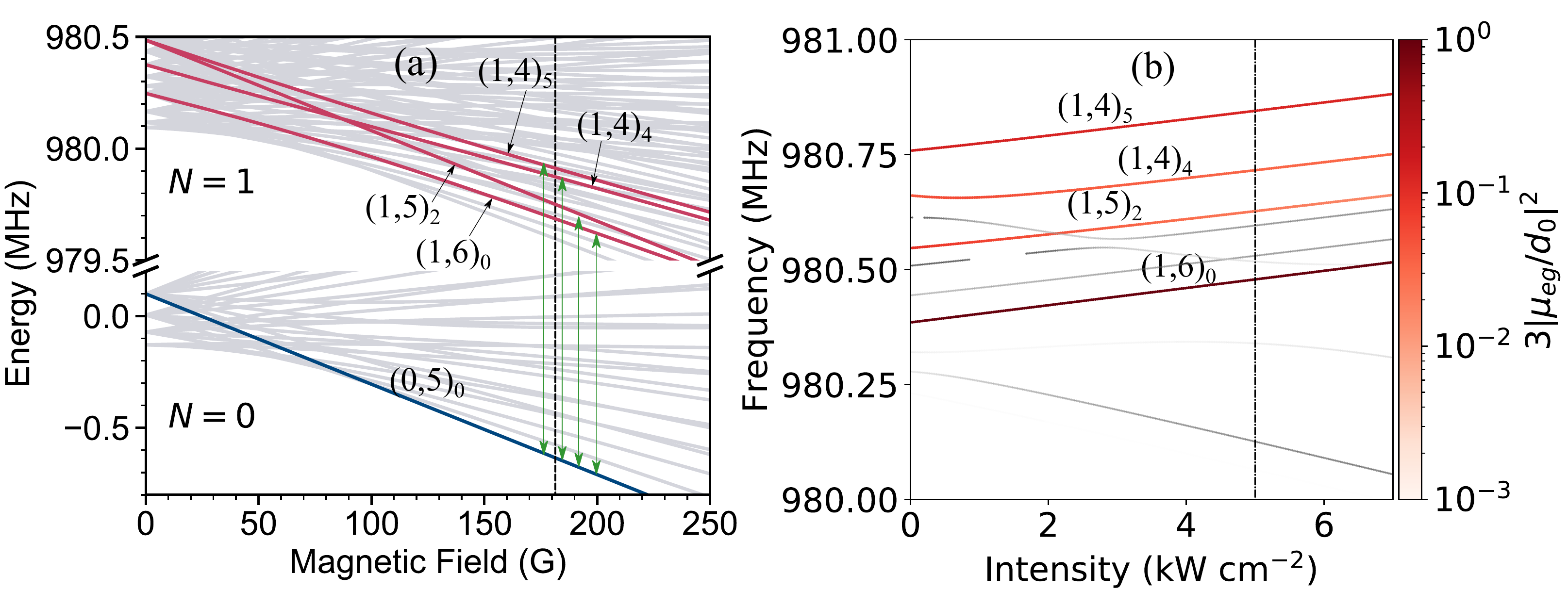}
	\centering
	\caption{(a) Zeeman splitting of hyperfine levels of $^{87}$Rb$^{133}$Cs. The vertical dashed line indicates a magnetic field $B=181.5$ G. Levels that form a 4-level qudit with $N=1$ are highlighted in red. The level in $N = 0$ that can be used to  manipulate them is highlighted in blue, with the transitions shown by the green arrows (separated in magnetic field for clarity). (b) Frequencies of transitions from level $(0,5)_0$, including the ac-Stark shift. Intensities are shown by shading. The dot-dashed line indicates an intensity $I_0=5$ kW cm$^{-2}$.} 
	\label{RbCs_levels_N1}
\end{figure}
The number of hyperfine levels in the $N = 1$ manifold is three times larger than for $N = 0$. 
This has two advantages: (i) higher-dimensional qudits can be formed; (ii) the ratio of the number of levels to the number of transitions is higher, making it more likely that there are isolated transitions. 
Using the search algorithm we find 21 hyperfine levels with $p^\textrm{max}_\textrm{loss} < 10^{-3}$ for $t_{\pi/2} = 1$~ms and 11 for $t_{\pi/2} = 0.3$ ms. However, a disadvantage of using levels with $N = 1$ is that they suffer decoherence due to differential ac-Stark shifts in the field of the trapping laser~\cite{Gregory2017}; this is discussed further in section~\ref{decoherence_sec}. We select four levels from the set for $t_{\pi/2} = 0.3$ ms that have a common $N=0$ level and minimal differential ac-Stark shifts ($\sim 1$ kHz/[kW cm$^{-2}$]). These are listed in table~\ref{rbcs_n1_levels} and highlighted in figure~\ref{RbCs_levels_N1}, together with the transitions that can be used to manipulate them. 
\begin{table}[h]
\begin{center}
    \begin{tabular}{||p{1.1cm}|p{1.1cm}|p{2cm}|p{1.8cm}||}
    \hline
    $e$  & $g$   & $f$/kHz & $3|\mu_{eg}|^2/d_0^2$\\
    \hline
    $(1,6)_0$ & $(0,5)_0$   & 980478.82	& 1\\
    $(1,5)_2$ & $(0,5)_0$   & 980627.29 & 0.02 \\
    $(1,4)_4$ & $(0,5)_0$   & 980716.34	& 0.03\\
    $(1,4)_5$ & $(0,5)_0$   & 980845.61	& 0.13\\
        \hline
    \end{tabular}
\end{center}
\caption{Levels $e=(N,m_F)_i$ that form a 4-level qudit for $^{87}$Rb$^{133}$Cs with $N=1$, together with the isolated transitions to levels with $N=0$ that can be used to manipulate them at a magnetic field of 181.5~G and tweezer intensity of 5 kW cm$^{-2}$.}
\label{rbcs_n1_levels}
\end{table}

\section{Sources of decoherence and gate errors}\label{decoherence_sec}
In this section, we discuss the main decoherence mechanisms for a qudit formed from a single ultracold molecule. 
For the states of $^{87}$Rb$^{133}$Cs considered here, the decoherence rate due to spontaneous emission and room-temperature blackbody radiation will be less than $10^{-5}$~Hz~\cite{Buhmann2008} and can be ignored. For $^{40}$Ca$^{19}$F the excitation from room temperature blackbody radiation results in a decoherence rate of $\sim 0.2$~Hz, which can be reduced to~$\sim 10^{-5}$~Hz at a temperature of 77~K~\cite{Buhmann2008}.

The non-deterministic variations in the energy differences between the levels that form the qudit will also cause decoherence. For isolated molecules, this non-deterministic variation can arise due to noise in electric, magnetic and electromagnetic fields. For uncorrelated white noise, the decay in coherence is exponential~\cite{Brouard2003}. If the standard deviation of the energy difference between a pair of states is $h\Delta\delta$, the coherence time is $\tau_\textrm{d} \sim 1/(\Delta\delta)$.

Ultracold molecules in optical lattices and tweezers are subject to ac-Stark effects. These are more complicated than for atoms because the molecular polarizability is anisotropic. For levels in the $N = 0$ manifold, there is no differential shift in first order, but there are small second-order shifts $\sim 1$ Hz~\cite{Park2017} that can lead to decoherence. A superposition of two $N = 0$ levels in $^{23}$Na$^{40}$K has been observed to retain its coherence for a time of around 1 second~\cite{Park2017} in an optical trap. For levels with $N \ge 1$, by contrast, there are first-order differential ac-Stark shifts due to the anisotropic part of the molecular dynamic polarizability, $\alpha^{(2)} = \frac{2}{3}(\alpha_{\parallel}-\alpha_{\bot})$~\cite{Gregory2017,Kotochigova2010}, where $\alpha_{\parallel}$ and $\alpha_{\bot}$ are the polarizabilities parallel and perpendicular to the internuclear axis.

In a thermal sample, variations in the laser intensity across the sample can lead to decoherence. For molecules in the motional ground state, however, only noise in the intensity can result in decoherence.  
In the worst case the coherence time will be roughly $h/(\alpha^{(2)} \Delta I)$, where $\Delta I$ is the noise in the intensity. For both $^{40}$Ca$^{19}$F and $^{87}$Rb$^{133}$Cs in $N=1$, $\Delta I \approx 1 \times 10^{-3}\,I_0$ gives a coherence time of at least $10$\,ms.
In practice, longer coherence times are possible because there are pairs of levels whose differential polarizability is much smaller than $\alpha^{(2)}$~\cite{blackmore_ultracold_2018}. For the four $N = 1$ levels of $^{40}$Ca$^{19}$F and $^{87}$Rb$^{133}$Cs molecules selected in sections~\ref{2sigma} and~\ref{1sigma}, we calculate coherence times of roughly 25 ms and 200 ms respectively.

The fluctuation in transition frequency $\Delta f$ due to intensity noise will also result in an error for the microwave gates described in section \ref{gate_sec} for both $N = 0$ and $N = 1$ levels. The upper bound for this error can be estimated as $\Delta f^2/\Omega^2 = (\alpha^{(2)} \Delta I/\hbar \Omega)^2 = (2t_{\pi/2}\alpha^{(2)} \Delta I/\pi\hbar)^2$. This gives an error of $10^{-4}$ for $^{40}$Ca$^{19}$F with $t_{\pi/2} = 5$~$\mu$s and $10^{-3}$ for $^{87}$Rb$^{133}$Cs with $t_{\pi/2} = 0.3$~ms.

The ac-Stark effect will also introduce differences in the trapping potentials for molecules with $N \ge 1$. This will lead to differences in the resonant frequencies for molecules in different motional states. However, recent experiments~\cite{Kaufman_2012,Thompson_2013,Liu_2019} have succeeded in cooling atoms to their motional ground-state with a probability of greater than 0.990. We believe that such techniques can be extended to molecules and will reduce gate errors to below 1\%. 

Another contribution to decoherence will be from noise in the magnetic field. The coherence time is $\tau_\textrm{d} \sim h/(\Delta B \Delta \mu )$, where $\Delta B$ is the standard deviation in the magnetic field and $\Delta \mu$ is the difference in the magnetic moments of the levels that form the qudit. It is relatively straightforward to achieve noise below 50~mG at fields of order $100$~G~\cite{barbe_observation_2018}. 
For rotational and hyperfine levels of $^{87}$Rb$^{133}$Cs, $\Delta \mu \sim  g_{n}\mu_\textrm{N}$, where $g_n$ is the nuclear g-factor and $\mu_\textrm{N}$ is the nuclear magneton. This gives a coherence time $\tau_\textrm{d} \sim 4$ s. 
For $N = 0$ levels of $^2\Sigma$ diatomic molecules, the coherence time can be estimated as $\tau_\textrm{d} \sim h/(g_e \mu_\textrm{B} \Delta B)$, where $g_e$ is the electron g-factor and $\mu_\textrm{B}$ is the Bohr magneton. For a magnetic field noise of 50 mG, $\tau_\textrm{d} \sim 10 \ \mu$s. However, with greater effort it is possible to reduce noise to below 50~$\mu$G~\cite{merkel_magnetic_2018,farolfi_design_2019}, resulting in a coherence time $\tau_\textrm{d} \sim 10$ ms. For the four levels of $^{40}$Ca$^{19}$F selected in section~\ref{2sigma}, $\Delta \mu \sim 10^{-2}g_e \mu_\textrm{B}$, giving a coherence time $\tau_\textrm{d} \sim 400~$ms under these conditions.
It may be possible to increase the coherence time further by using levels with nearly equal magnetic moments \cite{Caldwell_2019}.

Our proposal does not involve a static electric field. Linear molecules in $\Sigma$ states have quadratic dc-Stark effects, so decoherence due to electric field noise will be insignificant. 

The analysis above shows that, under appropriate experimental conditions, the qudits formed from ultracold molecules can have long coherence times compared to the gate duration. 

\section{Microwave gates for molecular qudits}\label{gate_sec}
The Hamiltonian for the interaction between the molecule and a microwave field is 
\begin{equation}
    \begin{array}{ll}
        H(t) &= H_0 +V(t) \\
        &= \sum_{i = 1}^n \hbar \omega_i \ket{i}\bra{i} + \sum_{i\ne j} \hbar \left(\frac{\Omega_{ij}}{2}e^{(-i \omega_{ij} t + i \phi_{ij})}\ket{i}\bra{j}+ \frac{\Omega_{ij}}{2}e^{(i \omega_{ij} t - i \phi_{ij})}\ket{j}\bra{i}\right),
    \end{array}
\end{equation}
where $\hbar \omega_i$ is the energy of level $i$, $\omega_{ij}$ is the frequency of a microwave field resonant with the transition $i \leftrightarrow j$ and $\Omega_{ij}$ is the Rabi frequency. $\phi_{ij}$ denotes the phase of the microwaves. 
Using the unitary transformation $U(t) =\sum_{i = 1}^n e^{- i \omega_i t} \ket{i}\bra{i}$, the Hamiltonian in the generalized rotating frame~\cite{Leuenberger_2004} becomes
\begin{equation}
H_\textrm{r} = U(t)H(t)U^{\dagger}(t) 
 +i\hbar U(t) \frac{\partial U^{\dagger}(t)}{\partial t} 
= \sum_{i\ne j} \hbar \left(\frac{\Omega_{ij}}{2}e^{ i \phi_{ij}}\ket{i}\bra{j} + \frac{\Omega_{ij}}{2}e^{- i \phi_{ij}}\ket{j}\bra{i}\right).
\end{equation}
In an experiment we can measure only probabilities. The operator for such a measurement is $M =\ket{i}\bra{i}$. In the rotating frame, $ U(t)MU^{\dagger}(t) = M$.

Consider the case where the microwaves address two hyperfine levels $k$ and $l$ with the same $N$ via a common level $c$ with $N-1$. The Rabi frequencies for the transitions $k\leftrightarrow c$ and $l\leftrightarrow c$ are $\Omega_{kc}$ and $\Omega_{lc}$ respectively. The unitary evolution operator after time $t$, $\exp[-i H_\textrm{r} t/\hbar]$, in this 3-level subspace (with basis $\ket{c}, \ket{k}$, $\ket{l}$) is
\begin{equation}\label{U0}
\begin{bmatrix}
 \cos[\tilde{\Omega} t / 2] & \frac{-i\Omega_{kc} e^{i \phi_{kc}}\sin[ \tilde{\Omega} t / 2]}{\tilde{\Omega}} & \frac{-i\Omega_{lc} e^{i \phi_{lc}} \sin[ \tilde{\Omega} t / 2]}{\tilde{\Omega}} \\
 \frac{-i\Omega_{kc} e^{-i \phi_{kc}} \sin[ \tilde{\Omega} t / 2]}{\tilde{\Omega}} & \frac{\Omega_{lc}^2+\Omega_{kc}^2\cos[ \tilde{\Omega} t / 2] }{\tilde{\Omega}^2} & \frac{e^{-i (\phi_{kc} - \phi_{lc})} \Omega_{kc} \Omega_{lc} \left(\cos[ \tilde{\Omega} t / 2]-1\right)}{\tilde{\Omega}^2} \\
 \frac{-i\Omega_{lc} e^{-i \phi_{lc}} \sin[ \tilde{\Omega} t / 2]}{\tilde{\Omega}} & \frac{e^{i (\phi_{kc} - \phi_{lc})} \Omega_{kc} \Omega_{lc} \left(\cos[ \tilde{\Omega} t / 2]-1\right)}{\tilde{\Omega}^2} & \frac{\Omega_{kc}^2+\Omega_{lc}^2\cos[ \tilde{\Omega} t / 2] }{\tilde{\Omega}^2}, 
\end{bmatrix},         
\end{equation}
where $\sqrt{\Omega_{kc}^2+\Omega_{lc}^2} = \tilde{\Omega}$.
We choose square pulses for the two microwave fields with pulse duration $2\pi/\tilde{\Omega}$, such that there is no population transfer to the common level $\ket{c}$.  For $\zeta = \Omega_{lc}/\Omega_{kc}$ and $\phi = \phi_{kc} - \phi_{lc}$, the operator \eqref{U0} becomes 
\begin{equation}\label{U1}
\mathscr{U}_{k,l}(\zeta,\phi) = 
    \begin{bmatrix}
     \frac{2}{\zeta^2+1}-1 & -\frac{2 \zeta e^{-i \phi}}{\zeta^2+1} \\
    -\frac{2 \zeta e^{i \phi}}{\zeta^2+1} & 1-\frac{2}{\zeta^2+1} \\
    \end{bmatrix},
\end{equation}
in the subspace $\{\ket{k},\ket{l}\}$. 
Using the ratio $\zeta$ and phase $\phi$, we can engineer gates between the hyperfine levels of the $N$ manifold. 

Similarly, we can create a phase gate for each of the levels in the $N$ manifold. For $\Omega_{lc} = 0$ and  $t = \pi/\Omega_{kc}$ equation~\eqref{U0} becomes
\begin{equation}\label{Rz0}
 Q_{k}(\phi_{kc}) =
\begin{bmatrix}
 0 & -i e^{i \phi_{kc}} & 0 \\
 -i e^{-i \phi_{kc}}  &0 &0 \\
0 & 0& 1
\end{bmatrix}.         
\end{equation}
A phase gate $R_k(\phi)$ can be created for state $k$ from two such operations,
\begin{equation}
R_k(\phi) = Q_{k}(\pi-\phi)  Q_{k}(0).
\label{eq:phase-gate}
\end{equation}
One of the advantages of using microwaves is that the phase $\phi$ can be controlled precisely. 
 
If the common state is a level of the $N + 1$ rotational manifold instead of $N-1$ as above, analogous gates can be obtained by substituting $\mathscr{U}_{k,l}(\zeta,2\pi-\phi)$ for $\mathscr{U}_{k,l}(\zeta,\phi)$ in the above equations.

\section{Quantum algorithm using a qudit}\label{Deutsch_section}
\begin{figure}[h]
	\includegraphics[width=10cm]{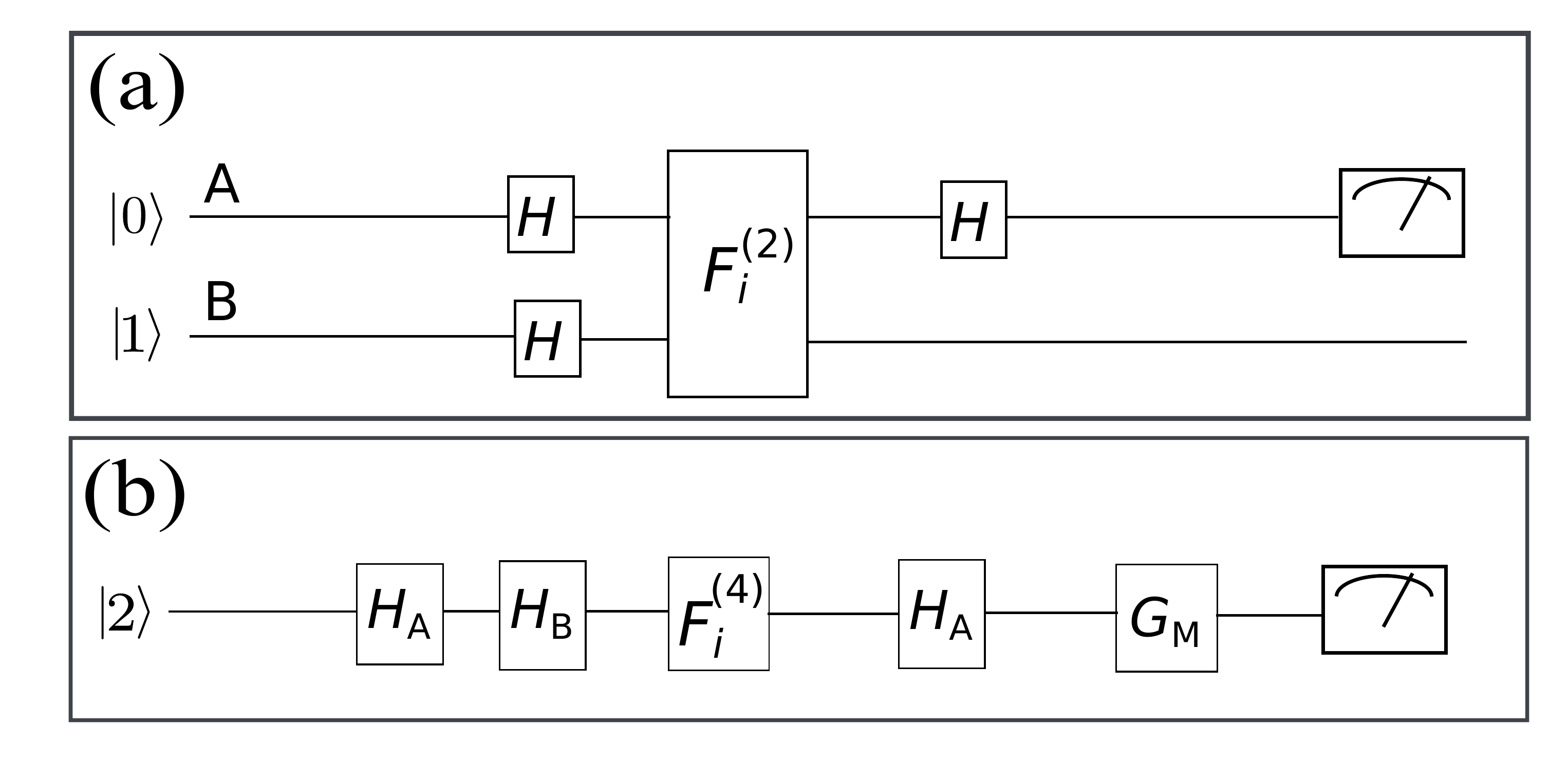}
	\centering
	\caption{Quantum gates for implementing the Deutsch algorithm using (a) two qubits and (b) a single 4-level qudit.  The gate operators for the qudit are given in equation~\eqref{state_mapping}.}
	\label{Deutsch}
\end{figure}
As a practical application of ultracold molecules as qudits, we propose an implementation of the Deutsch algorithm~\cite{Deutsch_david_quantum_1985}.
Consider the four possible one-bit Boolean functions,
\begin{equation}\label{boolean}
\begin{array}{ll}
        f_1(0) = 0, & f_1(1) = 0, \\
        f_2(0) = 1, & f_2(1) = 1, \\
        f_3(0) = 1, & f_3(1) = 0, \\
        f_4(0) = 0, & f_4(1) = 1.
    \end{array}
\end{equation}
The Deutsch algorithm determines whether a one-bit Boolean function $f$ is balanced (i.e., gives 1 for one input and 0 for the other) or constant (gives 0 for both inputs or 1 for both inputs).
Classical algorithms require at least two calls to the function $f$ to answer this question. 
Deutsch~\cite{Deutsch_david_quantum_1985} showed that an implementation with two qubits can answer the question with a single call to $f$.

Deutsch's implementation starts by initializing the two qubits in state $\ket{0}_\mathrm{A}\otimes\ket{1}_\mathrm{B}$, and subjects them to a series of gates, represented as a quantum circuit in figure~\ref{Deutsch}(a).
The operator $F_i^{(2)}$ in Fig.~\ref{Deutsch}(a) is a quantum implementation of the function $f_i$ that maps the two-qubit state $\ket{x}\ket{y}$
to $\ket{x}\ket{f_i(x)\oplus y}$, with $\oplus$ the sum modulo 2~\cite{Deutsch_david_quantum_1985,Kiktenko2015}, and $H$ is the Hadamard operator.
At the end of the circuit, the state of the first qubit is $\ket{1}_\mathrm{A}$ if $f_i$ is balanced and $\ket{0}_\mathrm{A}$ if it is constant~\cite{Deutsch_david_quantum_1985}.

Recently, Kiktenko \textit{et al.}~\cite{Kiktenko2015} proposed an alternative implementation using a qudit with 4 primary levels.  
As a first step, they map the two-qubit basis states onto a four-level qudit basis, $\{ \ket{j}, j=1,\ldots4 \}$, according to
\begin{equation}\label{state_mapping}
    \begin{array}{ll}
        \ket{0}_\mathrm{A}\otimes\ket{0}_\mathrm{B} \to \ket{1},
        & \ket{0}_\mathrm{A}\otimes\ket{1}_\mathrm{B} \to \ket{2}, \\
        \ket{1}_\mathrm{A}\otimes\ket{0}_\mathrm{B} \to \ket{3},
        & \ket{1}_\mathrm{A}\otimes\ket{1}_\mathrm{B} \to \ket{4}. 
    \end{array}
\end{equation}
The function of interest, $f_i$, is mapped onto a unitary operator $F_i^{(4)}$ that acts on the qudit space as described by equation (\ref{gates}) below.
To determine the character of $f_i$, the qudit is initialised in state $\ket{2}$ and subjected to the circuit in Fig.~\ref{Deutsch}(b) using the unitary operator $F_i^{(4)}$ associated with $f_i$. To find out whether the function is constant or balanced, it suffices to determine whether the molecule at the end of the evolution is in state $\ket{2}$ or not: If the molecule is in state $\ket{2}$, the function is constant; otherwise, it is balanced~\cite{Kiktenko2015}. Thus, after the MW pulses implementing the circuit, the molecule is subject to a projective measurement on state $\ket{2}$.  For $^{40}$Ca$^{19}$F, this state can be detected by laser-induced fluorescence (LIF)~\cite{Williams2018}. For $^{87}$Rb$^{133}$Cs, hyperfine-resolved STIRAP can be used to transfer the molecule to a Feshbach state that is then dissociated into constituent atoms~\cite{Gregory2017}. The atoms are then detected by LIF.

The circuit in Fig.~\ref{Deutsch}(b) involves the set of gates $\{H_\text{A}$, $H_\text{B}$, $G^\text{CNOT}_{\text{B} \rightarrow \text{A}}$, $G^\text{CNOT}_{\text{A} \rightarrow \text{B}}$, 
$F_1^{(4)}$, $F_2^{(4)}$, $F_3^{(4)}$, $F_4^{(4)}$, $G_\text{M}\}$ in the $d=4$ qudit space.
These
are defined
using the single-qubit 
gates from section~\ref{gate_sec} as
\begin{align}
    H_\text{A} &= \mathscr{U}_{1,3}(\sqrt{2}-1,\pi)\  \mathscr{U}_{2,4}(\sqrt{2}-1,\pi)\ ,
    & F_1^{(4)} & = I(4)\ , 
    \nonumber \\
    H_\text{B} &= \mathscr{U}_{1,2}(\sqrt{2}-1,\pi)\  \mathscr{U}_{3,4}(\sqrt{2}-1,\pi)\ ,
    & F_2^{(4)} & = \mathscr{U}_{1,2}(1,\pi)\  G^\text{CNOT}_{\text{A} \rightarrow \text{B}}\ ,
    \nonumber \\
    G^\text{CNOT}_{\text{A} \rightarrow \text{B}} &= \mathscr{U}_{3,4}(1,\pi)\ ,
    &
    F_3^{(4)} &= G^\text{CNOT}_{\text{A} \rightarrow \text{B}}\ , 
    \nonumber \\
    G^\text{CNOT}_{\text{B} \rightarrow \text{A}} & = \mathscr{U}_{2,4}(1,\pi)\ , 
    & F_4^{(4)} &= \mathscr{U}_{1,2}(1,\pi)\ ,
    \nonumber \\
    G_\text{M} &= \mathscr{U}_{1,2}(\sqrt{2}-1,\pi)\ .
    \label{gates}
\end{align}
Here $I(4)$ is the identity operator of dimension 4 and $\{\text{A,B}\}$ identify the qubit spaces. 
The operators $F_i^{(4)}$ acting on the qudit states $\ket{1},\ldots,\ket{4}$ result in the same states in the qudit space as the two-qubit operators $F_i^{(2)}$ acting on the states 
$\ket{0}_\text{A}\ket{0}_\text{B}$,
$\ket{0}_\text{A}\ket{1}_\text{B}$,
$\ket{1}_\text{A}\ket{0}_\text{B}$, and
$\ket{1}_\text{A}\ket{1}_\text{B}$; see Ref.~\cite{Kiktenko2015} for details.

This algorithm can be implemented using the four levels of the $^{40}$Ca$^{19}$F molecule identified in section~\ref{2sigma}, with the mapping $\ket{1, 1_u,-1} \to \ket{1}, \ \ket{1, 0,0} \to \ket{2},\ \ket{1, 1_l,1} \to \ket{3},\ \ket{1, 1_l,0} \to \ket{4}$. The total time required to apply all the gates will be $t_{\text{tot}} \approx 140\ \mu$s, assuming that the maximum Rabi frequency is $\pi/(2t_{\pi/2})$, with $t_{\pi/2} = 5 \ \mu$s. The error due to decoherence will then be $t_{\text{tot}}/\tau_\textrm{d} \sim 10^{-2}$. The total error due to off-resonant excitation from all gates, calculated using equation~\eqref{loss_frac}, will be $\sim 10^{-5}$. The total gate error due to the uncertainty in the frequency of the transition will be $10^{-3}$. This will result in a total error of only $\sim 10^{-2}$ in the computed output, without any error correction.    

The $N = 0$ hyperfine levels of ultracold $^{87}$Rb$^{133}$Cs molecules shown in table~\ref{rbcs_n0_levels} can also be used to define a $d = 4$ qudit space, with the mapping $\ket{0,3}_0 \to \ket{1}, \ \ket{0,5}_0 \to \ket{2},\ \ket{0,4}_0 \to \ket{3},\ \ket{0,4}_1 \to \ket{4}$. As there is no direct two-photon transition between levels 1 and 3, $H_\text{A}$ needs a longer sequence of gates, $H_\text{A} = \mathscr{U}_{1,2}(1,0)\ \mathscr{U}_{2,3}(\sqrt{2}-1,0)\ \mathscr{U}_{1,2}(1,0)\  \mathscr{U}_{2,4}(\sqrt{2}-1,\pi)$. For $t_{\pi/2} = 0.3$ ms, the total time required to apply all the gates will be $t_{\text{tot}} \approx 10$ ms. The error due to decoherence, off-resonant excitation, and frequency uncertainty will be $\sim 10^{-2}$, $10^{-2}$ and $10^{-2}$ respectively, giving a total error of only $\sim 10^{-2}$ in the computed output. For a qudit formed from $N=1$ levels of $^{87}$Rb$^{133}$Cs, the total error is around $5\times10^{-2}$, because of the additional decoherence from the ac-Stark effect.

\section{Conclusion}
We have examined the rich internal structure of $^{2}\Sigma$ and $^{1}\Sigma$ molecules, with a view to using the internal levels as qudits for quantum information processing using microwave pulses. We have analyzed two molecules of current experimental interest, $^{40}$Ca$^{19}$F and $^{87}$Rb$^{133}$Cs, confined in the motional ground states of optical tweezers. The large splitting between the hyperfine levels of $^2 \Sigma$ molecules compared to $^1 \Sigma$ molecules is advantageous in reducing off-resonant excitation of neighbouring levels. Nevertheless, we have identified possible implementations of 4-level qudits in both $^{40}$Ca$^{19}$F and $^{87}$Rb$^{133}$Cs, using a magnetic field to engineer suitable level spacings, transition strengths and field sensitivities. We have discussed two primary sources of decoherence for qudits formed from these levels: (i) differential ac-Stark shifts due to intensity noise in the trapping laser; (ii) magnetic field noise.  A major advantage of $^1\Sigma$ molecules is the very slow decoherence induced by magnetic field noise, which arises because their magnetic sensitivity is typically three orders of magnitude smaller than for $^2\Sigma$ molecules. Hyperfine levels with $N = 0$ have equal ac-Stark shifts for both molecules and are therefore very stable against decoherence associated with laser intensity noise. 

We have derived a set of gates, based on microwave transitions, for a qudit formed from a single ultracold molecule. We have shown how a sequence of microwave pulses applied to a polar molecule can be used to implement the Deutsch algorithm. 
Our calculations indicate that the algorithm can be executed in 0.14\,ms using $^{40}$Ca$^{19}$F and 10\,ms using $^{87}$Rb$^{133}$Cs, with an error $\sim 10^{-2}$ in each case. The Deutsch algorithm provides a proof-of-principle experiment to demonstrate the use of ultracold molecules to perform quantum computation. 
Scalability may be achieved in the future by implementing gates involving multiple molecules, confined in an array of tweezers and linked by the dipole-dipole interaction~\cite{DeMille2002,Yelin_2006,Zhu_2013,Herrera_2014,Karra_2016,Ni2018,hughes_robust_2019}.

\ack
We acknowledge stimulating discussions with Lincoln Carr.
This work was supported by U.K. Engineering and Physical Sciences Research Council (EPSRC) Grants
EP/P01058X/1, 
EP/P009565/1, 
EP/P008275/1, 
and EP/M027716/1. 
D. Jaksch acknowledges support from the European Research Council Synergy Grant Agreement No. 319286 Q-MAC.
J. Aldegunde acknowledges funding by the Spanish Ministry of Science and Innovation (grants MINECO/FEDER-CTQ2015-65033-P and PGC2018-096444-B-I00).

The data, code and analysis associated with this work are available at doi:10.15128/r1sn009x78c.

\bibliography{Qudits.bib}{}

\end{document}